\documentstyle[A4,12pt]{article}
\begin{document}
\newenvironment{abstr}
{\normalsize
 \quotation}
{\endquotation}
\title{{\Large \bf Aharonov-Bohm effect
in curved space and cosmic strings}}
\author{Yu.A.Sitenko \and A.V.Mishchenko \and
N.N.Bogolyubov Institute for Theoretical Physics \\
National Academy of Sciences of Ukraine  \\
252143, Kiev, Ukraine}
\date{ }
\maketitle
\begin{abstr}
\begin{sloppypar}
A quantum theory is developed for the scattering of a nonrelativistic
particle in the field of a cosmic string regarded as a combination
of a magnetic and gravitational strings. Allowance is made for the
effects due to the finite transverse dimensions of the string under
fairly general assumptions about the distribution of the magnetic
field and spatial curvature in the string. It is shown that in a
definite range of angles the differential cross section at all
absolute values of the wave vector of the incident particle depends
strongly on the magnetic flux of the string.
\end{sloppypar}
\end{abstr}
\thispagestyle{empty}
\newpage

\pagenumbering{arabic}

It is well known that in classical mechanics a complete description of
electromagnetic effects can be made by means of the electromagnetic field
strength, which acts directly on charged particles. In 1959, Aharonov
and Bohm \cite{AB1}, using the Schr\"odinger equation, considered the
scattering of an electron in an external static magnetic field
produced by an
infinitely long solenoid and found an effect that does not depend on the
depth of penetration of the electrons into the region of magnetic force
lines. This showed that in quantum mechanics the electromagnetic field
acts on charged particles even in the case when the region in which the
field is localized cannot be reached by the particles (see, for example,
the reviews of Refs. \cite{Peshkin} and \cite{Afanas'ev}).

In this paper, we investigate the effect of curvature of space on the
Aharonov-Bohm effect.  Namely, we consider the situation in which
there is not only a tube of magnetic force lines but also an
extrnal static cylindrically symmetric gravitational field with
symmetry axis of that coincides with the axis of the magnetic tube. It is
assumed, that at large distances from the symmetry axis the
space becomes locally flat and the region of spatial curvature
(the gravitational tube) may either coincide with the magnetic tube or
include it or, finally, be included in it. We pose and solve the problem
of the scattering of charged test partiále in a space with
gravitational and magnetic tubes.

It is appropriate to point out here that the simultaneous existance of
magnetic and gravitational tubes is rather typical of models with spontaneous
breaking of gauge symmetries. In such models, there arise vacuum structures
in the form of strings (for example, Abrikosov-Nielsen-Olesen
vortices \cite{Abrikosov},\cite{NO}),
which are characterized, on the one hand, by a certain flux
that one can reasonably call magnetic since it corresponds to the spontaneous
breaking of a gauge degree of freedom and, on the other hand, by the value
a condensate that spontaneously breaks a symmetry. To this condensate there
corresponds a uniform distribution of mass along the string axis, and this
is the source of the gravitational field of the type described above.
We note that this latter field is rather weak and it is apparently for this
reason that it has effectively escaped the attention of investigators. In
particular, to the best of our knowledge the processes of scattering of
partiáles by strings (vortices) usually take into account only the
presence of the magnetic component, and the presence of the gravitational
component is ignored (see, for example, Refs. \cite{Alford}
and \cite{Ma}). It is
to the correction of this, in our view, shortcoming that the present
paper is devoted. We shall show that it is  possible to take into account
gravitational components of a much more general form than those inherent
in strings in models with spontaneous breaking of gauge symmetries.

In what follows, we shall,for definiteness, consider strings,
that are cosmological objects --- so-called cosmic strings. According
to modern ideas, the early stages in the evolution of the universe were
characterized by high temperatures and a greater degree of symmetry than now,
and as the universe cooled there was a series of phase transitions
with spontaneous symmetry breaking \cite{Kirzhnits}.
The topological defects formed
as a result (monopoles, strings, and domain walls) may be
stable, and the mere fact of their existence leads to many important
consequences in cosmology \cite{Kibble}.
Particulary interesting cosmological
objects are strings, which, in particular, can play the role of
nucleation centers of galaxies and
gravitational lenses \cite{Vilenkin}.

We believe it is of interest to consider the situation in which there is
flux of magnetic field along a gravitational string and to study the
scattering of a test charged particle by such an object, in other words,
to consider the standart (magnetic) Aharonov-Bohm effect in a Fierz space.
This problem was actually posed for the
first time in Ref. \cite{Sitenko1},
in which some results were obtained; in particular attention was
drawn to the difference between the scattering for
spinor and scalar partiáles.  In this paper, we make a more
systematic and complete study of the problem. We determine the $S$ - matrix
for scattering in the case of combined singular magnetic and gravitational
strings, establish the connection between the $S$ - matrix and the
scattering amplitude in this and in the more general case of
nonsingular strings, and also take into account effects
due to the finite transverse dimensions of the strings.

More precisely, we consider the spacetime with metric
\begin {equation}
ds^2 = -c^2dt^2 + dz^2 + f(X,Y)(dX^2 + dY^2) \,
\label {zh96.3}
\end {equation}
where $X = r \cos\phi \;, Y = r \sin\phi$ and $f$ is the conformal factor
of the metric of the surface $Z = const$. This last surface has
Gaussian curvature
\begin {equation}
K (X, Y) = -(2f)^{-1} (\partial^2_X + \partial^2_Y) \, {\rm ln} \, f \;,
\label {zh96.4}
\end {equation}
which satisfies the condition
\begin {equation}
K(X, Y) = 0
\label {zh96.5}
\end {equation}
for $ r> r _ K $. With regard to the closure of the
region, $ r\leq r _ K $, we assume that here $ K (X, Y) $ is a
piecewise continuous function with singularities at isolated points or
on isolated lines that are integrable with respect to the measure
$f(X, Y)dXdY$. Then the total integrated curvature (in
units of $2\pi$) is given by
\begin {equation}
\Phi_ K = \frac {1} {2\pi}
\int\limits _ {r\leq r _ K} dXdYf (X, Y) K (X, Y)\;.
\label {zh96.6}
\end {equation}
As we showed in Ref. \cite{Sitenko2},
for $r> r_K$ (when condition (\ref{zh96.5})
is satisfied) the conformal factor takes the form
\begin{equation}
F(X,Y) = \left (\frac {r} {r _ 0} \right)^{-2\Phi _ K}.
\label {zh96.7}
\end {equation}
Thus, taking into account
(\ref{zh96.3}), we obtain a relationship between the total
curvature of the surface and the linear mass density of
cosmic string
\begin {equation}
\Phi_K = 4GMc^{-2}.
\label {zh96.8}
\end {equation}
In the case of a singular gravitational string,
we have $r_K \to 0$ and
\begin {equation}
K(X,Y) = 2\pi\Phi_K \frac{\delta (X) \delta(Y)}{f(X,Y)} \; \;.
\label{zh96.9}
\end{equation}

Along coordinate lines corresponding to the variable $z$, in the
spacetime (\ref {zh96.3}) a static magnetic field $B(X,Y)$
that satisfies the condition
\begin{equation}
B(X,Y) = 0
\label {zh96.10}
\end{equation}
for $r> r_B$ is directed. With regard to the closure of
the region, $r \leq r_B$, we assume that here $B(X, Y)$ is
a piecewise continuous function with singularities at isolated
points or on isolated lines that are integrable with respect
to the measure $f(X,Y)dXdY$. Then the total magnetic flux
(in London units $2\pi\frac {\textstyle {\hbar c}}{\textstyle {e}}$)
is given by
\begin{equation}
\Phi = \frac {e}{2 \pi \hbar c}
\int\limits_{r\leq r_B} dXdYf(X,Y) B(X,Y) \;,
\label{zh96.11}
\end {equation}
where $e$ is the coupling constant of the matter to the gauge minus
field ($-e$ is the charge of the test partiále). In the case of a
singular magnetic string, we have $r_B \to 0$ and
\begin{equation}
B (X,Y) = 2\pi\frac{\hbar c}{e} \Phi
\frac{\delta (X) \delta (Y)}{f(X,Y)} \; \;.
\label{zh96.12}
\end{equation}

In this paper, we study the quantum--mechanical scattering of
nonrelativistic test partiáles by a cosmic string. Since the motion
of the partiáles along the string axis $z$ is free, we can make a
restriction to considering the motion of partiáles on the
surface $z=const$ (for more details see,
for example, Ref. \cite{Sitenko1}).The
Schr\"odinger equation for the wave function describing the
stationary scattering state has the form
\begin{equation}
H\psi (X,Y) = \frac{\hbar^2 k^2}{2m} \psi(X,Y) \;,
\label {zh96.13}
\end{equation}
where $m$ and $k$ are, respectively, the mass of the partiále and
the absolute value of its wave vector. Under the condition of axial
symmetry of the magnetic field strength and
the Gaussian curvature,
\begin{equation}
\partial_\phi B(X,Y) = 0, \quad \partial_\phi K(X,Y) = 0 \;,
\label{zh96.14}
\end{equation}
we obtain for the Hamiltonian the expression
\begin{equation}
H = -\frac{\hbar^2}{2m} \left\{\frac{1}{r^2 f} [(r\partial_r)^2 +
(\partial_\phi -i\frac{e}{\hbar c}V + \frac{1}{2}i\sigma W)^2] +
\sigma\frac{e}{\hbar c}B - \frac{1}{2} K\right\},
\label{zh96.15}
\end{equation}
where
\begin{equation}
V (r) = \int\limits_{0}^{r} drrfB, \quad
W (r) = \int\limits_{0}^{r} drrfK \;,
\label{zh96.16}
\end{equation}
$\sigma = \sigma_3 = \left(
\begin{array}{cc}
1&0 \\
0&-1
\end {array}
\right)$
in the case of a spinor partiále ($\psi$ is a two-component
column function) and $\sigma = 0$ in case of a scalar partiále
($ \psi $ is a single-component function). As a radial variable, it is
convenient to use the geodesic length in the radial direction:
\begin{equation}
\rho = \int\limits_{0}^{r} drf^{1/2} \;.
\label{zh96.17}
\end{equation}
Then the metric (\ref{zh96.3}) (under the condition of axial symmetry)
takes the form
\begin{equation}
ds^2 = -c^2dt^2 + dz ^ 2 + d\rho^2 + \gamma^2d\phi^2 \,,
\label{zh96.18}
\end{equation}
where
\begin{equation}
\gamma (\rho) = r (\rho) \sqrt{f[r(\rho)]} \;,
\label{zh96.19}
\end{equation}
and the Hamiltonian (\ref{zh96.15}) can be represented as follows:
\begin{eqnarray}
H = -\frac{\hbar^2}{2m} \left\{\gamma^{-1} \partial_\rho
\gamma\partial_\rho + \gamma^{-2} [\partial_\phi - i\frac{e}{\hbar c} V +
\frac{1}{2} i\sigma (1 - (\partial_\rho\gamma))]^2 +
\right. \nonumber \\
\left.
+ \sigma \frac{e}{\hbar c} \gamma^{-1} (\partial_\rho V) +
\frac{1}{2} \gamma^{-1} (\partial^2_\rho\gamma) \right \}.
\label{zh96.20}
\end{eqnarray}

The operator $H$ acts in the space of functions with  scalar product
\begin{equation}
(\psi, \psi \, ') =\int\limits_{0}^{2\pi}
d\phi\int\limits_{0}^{\infty} drrf\psi^{\ast} \psi \, ' =
\int\limits_{0}^{2\pi} d\phi\int\limits_{0}^{\infty} d\rho\gamma
\psi^{\ast} \psi \, ' \;.
\label{zh96.21}
\end{equation}
In contrast, the Hamiltonian in the absence of a string
(i.e. with $\Phi = \Phi_K = 0$),
\begin{equation}
H_0 = -\frac{\hbar^2}{2m} (\partial^2_\rho + \rho^{-1} \partial_\rho +
\rho^{-2} \partial^2_\phi) \;,
\label{zh96.22}
\end{equation}
acts in the space of functions with scalar product
\begin{equation}
(\psi, \psi \, ') =\int\limits_{0}^{2\pi} d\phi
\int\limits_{0}^{\infty} d\rho\rho\psi^{\ast} \psi \, ' \;.
\label{zh96.23}
\end{equation}
We introduce a transformation of the operator and of the functions
that leaves the form of equation (\ref {zh96.13}) unchanged
\begin{equation}
\tilde{H} = \Xi H \Xi^{-1}, \quad
\tilde{\psi} = \Xi\psi, \quad
\tilde{H} \tilde{\psi} = \frac{\hbar^2 k^2}{2m} \tilde{\psi} \;.
\label{zh96.24}
\end{equation}
Choosing
\begin{equation}
\Xi = \gamma^{1/2} \rho^{-1/2} \;,
\label{zh96.25}
\end{equation}
we obtain for the transformed Hamiltonian the expression
\begin{eqnarray}
\tilde{H}= -\frac{\hbar^2}{2m} \left\{ \partial^2_\rho + \rho^{-1}
\partial_\rho - \frac{1}{4} \rho^{-2} + \frac{1}{4} \gamma^{-2}
(\partial_\rho\gamma)^2 + \right. \nonumber \\
\left.
+ \gamma^{-2}
\left[ \partial_\phi - i\frac{e}{\hbar c} V + \frac{1}{2} i\sigma (1 -
(\partial_\rho\gamma)) \right]^2 + \sigma\frac{e}{\hbar c}
\gamma^{-1} (\partial_\rho V) \right\};
\label{zh96.26}
\end{eqnarray}
then in the space of the transformed functions the scalar product is defined
in accordance with (\ref{zh96.23}), rather than (\ref{zh96.21}).

We regard $H_0$ (\ref{zh96.22}) as an unperturbed Hamiltonian and
\begin{equation}
\Delta H = \tilde{H} -H_0
\label{zh96.27}
\end{equation}
as an operator that describes a perturbing interaction, and we shall
attempt to construct a scattering theory. Taking into account
(\ref{zh96.22}) and (\ref{zh96.26}), we  represent the operator
(\ref{zh96.27}) in the form
\begin{equation}
\Delta H = v (\vec x) + v^j (\vec x)
(-i\frac{\partial}{\partial x^j}) + v^{jj'} (\vec x)
( -\frac{\partial^2}{\partial x^j \partial x^{j'}}) \;,
\label{zh96.28}
\end{equation}
where we have introced the notation $\vec x = (x^1,x^2)\;,
\quad x^1 = \rho \cos\phi \;,\quad x^2 = \rho \sin\phi$ \\
and $j,j' = 1,2$.

If the coefficient functions $ v, \quad v^j$ and $v^{j,j'}$ decrease
as $O (\rho^{-1-\epsilon}) \quad (\epsilon > 0)$ as $\rho \to \infty$,
then in accordance with Ref. \cite{Hormander}
the perturbation $\Delta H$
(\ref{zh96.28}) has a short-range and scattering theory
can be constructed in the usual manner (see, for example
Refs. \cite{AGSitenko} and \cite{ReedSimon}). However,
even for partiále scattering by a purely
magnetic string ($\Phi \neq 0$ and $\Phi_K = 0$) the perturbation
$\Delta H$ has a long-range, since the coefficient function
$v^j$ decreases in the limit $\rho \to \infty$ as $O (\rho^{-1})$.
Because of the long-range interaction, it is impossible to choose a plane
wave as the incident wave, as noted by Aharonov and Bohm \cite{AB1}.
Nevertheless, in this case it is possible to develop a scattering
theory and obtain in its framework the Aharonov-Bohm scattering
amplitude (see Ref. \cite{Ruijsenaars}).

H\"ormander \cite{Hormander} considered a
class of perturbations of the form
(\ref{zh96.28}) containing  both a short-range part and a
long-range part characterized by real coefficients functions that
decrease in the limit $\rho \to \infty$ as
$O (\rho^{-\epsilon}) \quad (0 < \epsilon \leq 1)$, and he formulated
certain additional requirements under which scattering theory can be
developed. As he notes on page 417 of the translation of his monograph
of Ref. \cite{Hormander}
"the existence of modified wave operators is proved under
the weakest sufficient conditions among all those known at the present
time."

H\"ormander's conditions are satisfied by the perturbation in
the problema of scattering by a purely magnetic string,
\begin{equation}
v \sim O (\rho^{-2}) \mbox{    and    } v^j \sim O (\rho^{-1}),
\quad \rho \to \infty
\end{equation}
($v$ and $v^j$ are real functions, and $v^{jj'} = 0$), and, for
example, perturbation in the problem of scattering by a Coulomb center,
\begin{equation}
v \sim O (\rho^{-1}), \quad \rho \to \infty
\end{equation}
($v$ is a real function, and $v^j = v^{jj'} = 0$). In contrast,
the perturbation in the problem of scattering by a cosmic string
($\Phi \neq 0$ and $\Phi_K \neq 0$) does not satisfy the H\"ormander's
conditions:
\begin {equation}
v \sim O (\rho^{-2}) \, \quad v^j \sim O (\rho ^ {-1})
\mbox {    and    } v^{jj'} \sim O(1) \, \quad \rho \to \infty,
\label{zh96.29}
\end{equation}
where $v^j$, in contrast to $v$ and $v^{jj'}$ is a complex function
(more precisely, imaginary part of $v^j$ of order $\rho^{-1}$
is due to the nondecrease of the real quantity $v^{jj'}$
in the limit $\rho \to \infty$). Nevertheless, even in this last case it
is possible to develop a scattering theory, and in the
present paper (see Ref. \cite{paper})
we shall construct wave operators explicitly.

In Sec. 2, we determine the $S$-matrix and the scattering in
the case of a singular cosmic string ($r_B = 0$ and $r_K = 0$).
In Sec. 3, we take into account the effects of the gravitational
structure of the string (the finiteness of the transverse dimensions
of the region of curvature of space, $r_K> 0$), and in Sec. 4
the effects of the magnetic structure of the string
(finiteness of the trasverse dimensions of the region
of magnetic flux, $r_B> 0$). Sec. 5 is devoted to
discussion of the results. Details of the derivation of the
basic relations are given in Appendices A, B and C.

\vspace{.3in}
\noindent
{\bf Acknowledgments}\\
The work of Yu.A.S. and A.V.M. was supported by Swiss National Science
Foundation Grant No. CEEC/NIS/96-98/7 IP 051219.

\vfill
\eject

\end{document}